# Denial of Service Attack in Cooperative Networks

Tauseef Jamal, Zeeshan Haider, Shariq Aziz Butt, Assim Chohan
Email: [Tauseef.pieas, luckier19, shariq2315, asim2k994]@gmail.com

*Pakistan Institute of Engineering and Applied Sciences (PIEAS)*

Abstract--- In Denial of Service (DoS) attack the network resources are either delayed or refused to be assigned to the requested user [1]. This may occurs due to verity of reasons, could be intentionally or unintentionally. The unintentional case is quite hard to mitigate. In this paper we will refer the former case in context of cooperative networks. In cooperative networks we relay the data via intermediate nodes called relays. The relay selection is mechanism [2] need to be devised with focus on mitigating such attacks. In this paper we will enhance the relay selection mechanism address by [3] to propose the novel relay selection with emphasis on security of Wireless Local Area Networks.

## 1. INTRODUCTION

Cooperative networking is promising technique to mitigate the effect of fading in wireless networks. Due to cooperation we can achieve higher diversity, where the signal travels via multiple relay nodes. Cooperative communications have two parts, relay selection and data forwarding [5]. There are various relay selection mechanisms such as proactive relay selection, reactive relay selection [6], opportunistic relay selection [7] etc.

In proactive relay selections the relays are selected prior to transmission, while in reactive relaying the relays are selected when needed. Reactive relaying is beneficial when there is failed transmission. In case of opportunistic relay selections, the relays are selected on demand [8].

RelaySpot [9] is hybrid relaying protocol based on opportunistic relay selection mechanism. In RelaySpot the relays are selected based on history and interference, without taking into account of security. In the next section we explain how we incorporate ReValidation Mitigation Mechanism [10] in relay selection process to ensure secure relay selection.

Figure 1 shows the basic cooperation mechanism.

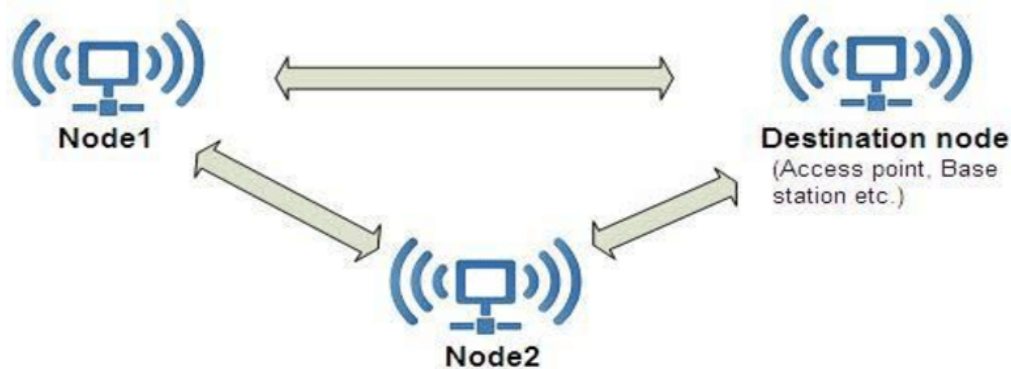

*Figure 1: Basic Cooperative Mechanism [14].*

Figure 2 demonstrate how we can achieve better rate due to cooperation. Since 802.11 rate adaptation allows the stations to adopt different rates based on distance and Signal to Noise Ratio (SNR). Therefore, the node far from AP would send data on lower data rate. As the Figure 2 shows that in presence of nodes at fast bit rate can overhear the data but they are obliged to drop it. But due to cooperation the intermediate node cooperate and act as relay to forward data on behalf of source. This way the medium is released earlier.

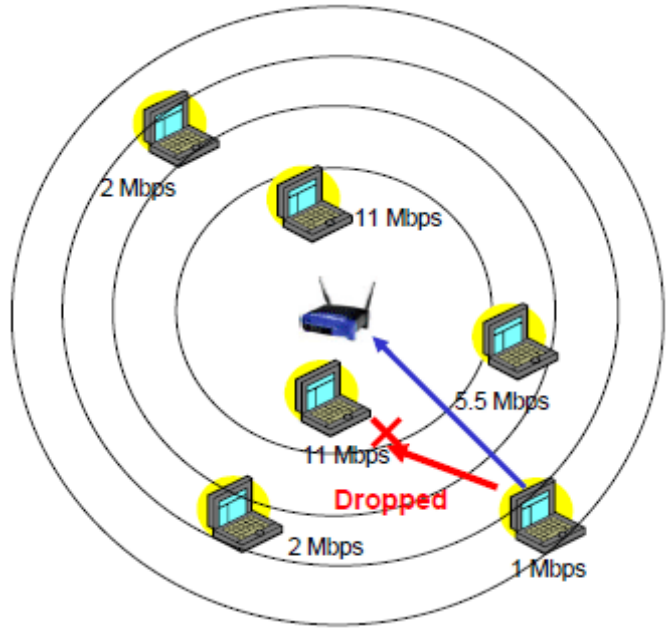

*Figure 2:802.11 Rate Adaptation [14].*

## 2. PROPOSED METHODOLOGY

DoS attack has many types, however in context of WLAN the most problematic is RTS attack. In this attack, the malicious node intentionally sends RTS control frame to Access Point (AP) or receiver keeping the AP or receiver unavailable for other nodes. On one hand it can leads to flooding but on the other hand such node send long reservation duration within field, which causes other nodes to wait for long. This loop hole in IEEE 802.11 STD has been addressed by [1] and [10]. In [10] T. Jamal et.al proposed the mitigation mechanism of such attack called Revalidation scheme.

### 2.1. Revalidation Scheme to RelaySpot

In RelaySpot the relays are selected based on history and interference information. History is computed by number of successful transmission while interference is computed by number of neighbors and number of simultaneous transmissions. However security has neither been considered during relay selection mechanism nor at cooperative transmission. In this work we are securing the both phases, however in this extended abstract we apply the Revalidation mechanism to the network and observed its effect.

The revalidation mechanism actually recomputed the reservation duration at AP, this additional check can verify if the sender of RTS is legitimate user or malicious. If AP found the sender as attacker, it immediately broadcast its MAC address to network so that no one can rely on such node and such node shouldn't be selected as relay node.

The problem with this mechanism is its complexity of computing and overhead of broadcasting. Therefore, we are working on more lightweight reliable solution. This first attempt just proves that security need to be addressed in cooperative relaying.

## 2.2. Relay Selection Mechanism

In literature we have various relay selection mechanism but most of them are relying over Channel State Information (CSI) computation [13]. CSI is not only hard to estimate but also not very reliable. In contrast to the prior art RelaySpot relies over local and stable information. Such that, it uses History Factor (HF) and Interference Factor (IF) to compute the Selection Factor (SF). Equation 1 shows the SF computation:

$$SF = HF / (1 + IF) , S \in ]0, 1] \qquad Eq\ (1)$$

HF is directly proportional to the history of successful transmission while IF is directly proportional to the processing delay of each node and node density. SF is computed locally and it guarantees the availability of node as well as fast reaction to the changes in network due to mobility and fading.

## 2.3. Request To Send (RTS) Attack

RTS attack is type of DoS attack where the sender of RTS sends RTS either to flood or to reserve the medium for undue amount of time. We are referring to the former case in our work. That is why we are using the revalidation technique.

In IEEE 802.11 standard the Medium Access Control (MAC) layer defines the channel access mechanism. It includes optional RTS/CTC handshake mechanism to avoid collision and hidden node problems. When the sender finds medium free it sends RTS frame after Distributed Inter Frame Space (DIFS) amount of time. DIFS is normally 50 micro seconds, and have considered the same in our evaluation. RTS frame includes 2 Bytes duration field to insert the duration of channel reservation for data transfer. This allows the overhearing nodes to remain in Quiet state for reservation duration. All other nodes will access the channel after expiration of this time. In case of malicious node the duration field in RTS is filled to undue amount of time such as maximum allowed time, causing the RTS attack.

## 3. EVALUATIONS

The implementations of RelaySpot are available in MiXim simulator of OMNet++ framework [11] and [12]. We have considered a scenario with one AP and several source nodes increasing from 5 to 50 nodes. Therefore, we implemented the Revalidation mechanism and run the simulations for 500 seconds within playground size of 500x500 meters. The payload size was assumed to be 2K bytes. Throughput has been computed at MAC instead of application layer. The results show that we have achieved performance gain of 40% in the presence of one malicious node. The same gain has been observed in the presence of multiple malicious nodes.

We have run the simulations for 50 times to achieve the confidence interval of 95%.

## 4. CONCLUSIONS AND FUTURE WORK

In this paper we have applied the DoS attack prevention mechanism to a cooperative networks. The results prove that cooperative network performed better in term of throughput in case of one or more malicious relay nodes. Which proved that DoS attack mitigation can outperform the cooperative communications.

As this is our work in progress, our aim is to apply security mechanism in relay selection mechanism, and cooperative transmission. This way we will enhance the RelaySpot to secure relaying protocols. We have just addressed the intentional attackers, as a future work we aim to address the unintentional attacks such as selfishness, unfairness and laziness.